\documentclass{IEEEtran}
\ifCLASSINFOpdf
 \usepackage[pdftex]{graphicx}
\else
\fi
\usepackage{cite}
\usepackage{amsmath,amssymb,amsfonts}
\usepackage{algorithmic}
\usepackage{graphicx}
\usepackage{textcomp}
\usepackage{subfigure}
\usepackage{threeparttable}
\usepackage{float}
\usepackage{epsfig}
\usepackage{epstopdf}
\usepackage{amsmath}
\usepackage{multirow}
\usepackage{color, soul}
\soulregister\cite7 % 针对\cite命令
\soulregister\citep7 % 针对\citep命令
\soulregister\citet7 % 针对\citet命令
\soulregister\ref7 % 针对\ref命令
\soulregister\pageref7 % 针对\pageref命令

\def\BibTeX{{\rm B\kern-.05em{\sc i\kern-.025em b}\kern-.08em
    T\kern-.1667em\lower.7ex\hbox{E}\kern-.125emX}}

\begin{document}

\title{\LARGE{Coverage Probability Analysis of IRS-Aided
Communication Systems}}

 \author{Zhuangzhuang~Cui,~\IEEEmembership{Student Member,~IEEE,}~Ke~Guan,~\IEEEmembership{Senior Member,~IEEE,}\\Jiayi~Zhang,~\IEEEmembership{Senior Member,~IEEE}~and~Zhangdui~Zhong,~\IEEEmembership{Senior Member,~IEEE}% <-this % stops a space

 \thanks{Manuscript received xx xx, 2020; revised xx xx, 2020; accepted xx xx, 2020. This work was supported by Key-Area Research and Development Program of Guangdong Province, China (2019B010157001), NSFC under Grant (61771036, 61901029), the State Key Laboratory of Rail Traffic Control and Safety (Contract No. RCS2020ZZ005). (\emph{Corresponding author: Ke Guan}).}
 \thanks{All the authors are with the State Key Lab of Rail Traffic Control and Safety, Beijing Jiaotong University, Beijing, 100044 China. (E-mail: \{cuizhuangzhuang, kguan, zhangjiayi, zhdzhong\}@bjtu.edu.cn.)}}

\markboth{}{Z.~Cui \MakeLowercase{\textit{et al.}}: Coverage Probability Analysis of IRS-Aided Communication Systems}

\maketitle
\begin{abstract}
The intelligent reflective surface (IRS) technology has received many interests in recent years, thanks to its potential uses in future wireless communications, in which one of the promising use cases is to widen coverage, especially in the line-of-sight-blocked scenarios. Therefore, it is critical to analyze the corresponding coverage probability of IRS-aided communication systems. To our best knowledge, however, previous works focusing on this issue are very limited. In this paper, we analyze the coverage probability under the Rayleigh fading channel, taking the number and size of the array elements into consideration. We first derive the exact closed-form of coverage probability for the unit element. Afterwards, with the method of moment matching, the approximation of the coverage probability can be formulated as the ratio of upper incomplete Gamma function and Gamma function, allowing an arbitrary number of elements. Finally, we comprehensively evaluate the impacts of essential factors on the coverage probability, such as the coefficient of fading channel, the number and size of the element, and the angle of incidence. Overall, the paper provides a succinct and general expression of coverage probability, which can be helpful in the performance evaluation and practical implementation of the IRS.
\end{abstract}

\begin{IEEEkeywords}
Coverage probability, channel condition, intelligent reflective surface, Rayleigh fading, signal-to-noise ratio.
\end{IEEEkeywords}
\section{Introduction}
\IEEEPARstart{I}{NTELLIGENT} reflective surfaces (IRSs) have attracted much attention in both academy and industry since they are regarded as potentials to smartly \emph{design} propagation environment by adjusting the phase and magnitude of reflective wave operated by a central controller, and thereby improve the signal quality \cite{b1}. In the IRS-aided communication system, IRS is expected to {\it passively reflect} transmitted signal, and then the resultant signal can be received in any desired direction. Thus, the IRS technology can remedy the defect of limited coverage in millimeter-wave communications \cite{b2}. Moreover, in the scenario where the line-of-sight (LOS) path is obstructed, the IRS can significantly enhance the received power strength through the constructive superposition of impinging waves from each element in the IRS. It is worth mentioning that the IRS does not need radio frequency (RF) links, which essentially differs from the traditional relay technology. In brief, the advantages of low-cost, multi-use and easy-to-deployment have impelled many researches to show solicitude for the IRS technology \cite{b3,b4,b11,b5,b6,b7,b8,b9}.

%\subsection{Motivation} %, which can facilitate
 %For IRS-assisted communications, there are many open issues, involving channel modeling, performance analysis, system design and optimization, among which
 In this paper, we aim at analyzing the coverage probability for IRS-assisted communications. To our best knowledge, very few studies comprehensively focus on this issue. Technical motivations are inspired by many challenges led by the new features of IRS \cite{b4}, summarized as follows. First, the performance analysis of \emph{large-size} IRS needs considering a new channel model that takes the physical size of surfaces into account \cite{b11}. Then, the amplitude of the \emph{composite channel} is represented as the sum of the product of two random variables (RVs) following specific distribution, which is difficult to determine the distribution of signal-to-noise ratio (SNR). Last but not least, due to many factors, such as the physical size of IRS and channel conditions, affecting the coverage probability, a comprehensive analysis is urgently needed.
 %(i) The feature of massive array elements with considerable size requires the channel model that takes the physical size and number of surfaces into consideration. (ii) Since each element in IRS reflects the signal from transmitter to the receiver, the total amplitude of channel is expressed as the sum of the product of two random variables (RVs) following specific distribution, which is intractable for obtaining the distribution of signal-to-noise ratio (SNR). (iii) Since many factors can affect the performance limits, an accurate and comprehensive analysis is needed.

Since most of studies focus on the optimal phase shift or beamforming design of IRS \cite{b5}, related works that put efforts on the coverage probability analysis of IRS-aided communication systems are relatively few. To obtain the analytical expression, the authors in \cite{b6, b7} used the central limit theorem (CLT) to simplify the resultant amplitude of channels from $N$ elements to a Gaussian distributed RV by assuming $N\gg1$. In \cite{b8}, the IRS-based composite channel was presented to be equivalent to a direct channel with Nakagami-\emph{m} fading, which yields that the instantaneous SNR follows Gamma distribution for a large $N$. In \cite{b9}, since non-reciprocal channels were assumed to be independent and identically distributed (i.i.d.) complex Gaussian, the obtained tractable form of SNR can be further used in analyzing the symbol error probability. In the above works, the large-scale channel model is merely assumed as square attenuation of distance, which overlooks the impact of the physical size of the IRS on the performance. Moreover, most of the works assume that the number of elements in the IRS is enough large so as to employ the CLT, nevertheless, which may not be in line with the actual situation.

Consider the above deficiencies, we incorporate a more realistic channel model and an arbitrary number of elements into the coverage probability analysis. The main contributions are summarized as follows: (i) We employ a \emph{realistic path loss model} that takes the physical size of IRS and angle of incidence into account, which ensures that we can quantify their impacts on coverage probability. (ii) For a \emph{single element}, we determine the distribution of resultant channel gain under Rayleigh fading, and then the exact closed-form of coverage probability is derived. (iii) For \emph{multiple elements}, we adopt Gamma distribution to approximate the obtained distribution of resultant channel by the method of moment matching and then a tractable expression of coverage probability is obtained. (iv) We comprehensively analyze many factors that can influence the coverage probability by numerous simulations. The results are useful in the design of practical IRS.

%We also explore how many elements in IRS can guarantee the complete coverage for desired users, herein we obtain numerical solutions, presenting good agreements with simulations.

%The reminder of the paper is organized as follows. Section~II introduces the system and channel models. Section~III analyzes the coverage probability, where the quasi-closed form is derived. Numerical simulations are conducted in Section IV to verify the correctness of theoretical derivation and show the influences of many factors. Section~V concludes this paper.

\section{System Model}
We consider a general IRS-aided communication system, where the direct link is nonexistent due to blockage, as shown in Fig.~1. As a result, the IRS with the size of $u\times w$ and the number of $N$ metallic elements reflects the wave from the source (S) to the destination (D). Besides, distances between the source and IRS (SR), the IRS and the destination (RD) are denoted as $d_s$ and $d_r$, respectively. It is worth noting that we consider the far-field transmission for the transceivers, which means that $d_s, d_r \ge \frac{2(\max{\{u,w\}})^2}{\lambda}$ where $\lambda$ is the wavelength. Moreover, the angle of incidence is denoted as $\theta_s$ that is the angle with the normal direction of the plane of the IRS.

\subsection{Channel Model}
Prior works focus on modeling the path loss for the IRS-assisted communications \cite{b10,b12}. For instance, in \cite{b10}, the authors proposed the free space path loss models for both in the near-field and far-field conditions. Moreover, experimental measurements were conducted to verify these models. In this paper, we intend to introduce the size and number of elements in the IRS into the path loss in the far field, thus, a realistic model is used as \cite{b12}
\begin{equation}
\label{pl}
 P_l(d_s,d_r,\theta_s)=\frac{G_s G_r}{(4\pi)^2}\left(\frac{uw}{d_s d_r}\right)^2\cos(\theta_s),
\end{equation}
where $G_s$ and $G_r$ are the antenna gains of transmitter and receiver, respectively.

Then, we denote the amplitudes of small-scale fading of SR link and RD link as $\alpha_i$ and $\beta_i$, where $i$ is the index of the element in the IRS. They both follow Rayleigh distribution as \cite{b6}, with probability density function (PDF) given by
\begin{equation}
f_{\alpha}(\alpha)=\frac{\alpha}{\sigma^2}\exp\left(-\frac{\alpha^2}{2\sigma^2}\right),
\end{equation}
where $\sigma$ represents the fading coefficient of the channel.

\subsection{Signal-to-Noise Ratio}
Assuming that channels of SR and RD links experience quasi-static Rayleigh fading without interference, the received signal at D can be expressed as
\begin{equation}
\label{receive}
y=\sqrt{P_sP_l}\left[\sum_{i=1}^{N}h_i\chi_i g_i\right]x+n_0,
\end{equation}
where $x$ is the transmit signal with power of $P_s$ and $\chi_i=\varrho_i(\phi_i)e^{j\phi_i}$ is the reflected coefficient produced by the $i$th element of the IRS, with $\varrho_i(\phi_i)$ = 1 for the ideal condition (${\forall}i = 1, 2, . . . , N$). Note that $h_i=\alpha_ie^{-j\vartheta_i}$ and $g_i=\beta_ie^{-j\varphi_i}$ are channel gains where we assume that the channel state information (CSI) knowledge is available \cite{b7}. Moreover, $n_0$ is the additive white Gaussian noise following $\mathcal{N}(0,\sigma_n^2)$.

Then, the received SNR can be expressed as
\begin{equation}
\gamma=\frac{\left|\sum_{i=1}^{N}\alpha_i\beta_ie^{j(\phi_i-\vartheta_i-\varphi_i)}\right|^2 P_s G_s G_r(uw\cos(\theta_s))^2}{(4\pi \sigma_n d_sd_r)^2}.
\end{equation}

In order to obtain the maximum value of $\gamma$, we need to optimally design the phase shit of each element. Following the method in \cite{b6}, by setting the phases $\phi_i=\vartheta_i+\varphi_i$, the maximum $\gamma$ can be obtained as
\begin{equation}
\gamma=\frac{A^2\bar{\gamma}}{d_s^2d_r^2},
\end{equation}
where $A=\sum_{i=1}^{N}\alpha_i\beta_i$. Moreover, the average SNR can be written as $\bar{\gamma}=\frac{P_s G_s G_r (uw\cos(\theta_s))^2}{16\pi^2 \sigma_n^2}$. It intuitively shows that the large size of the IRS can bring large average SNR where the term of $uw\cos(\theta_s)$ represents the total effective area of the beam on the IRS as seen from the source.

\begin{figure}[t!]
  \centering
\includegraphics[width=2.5in]{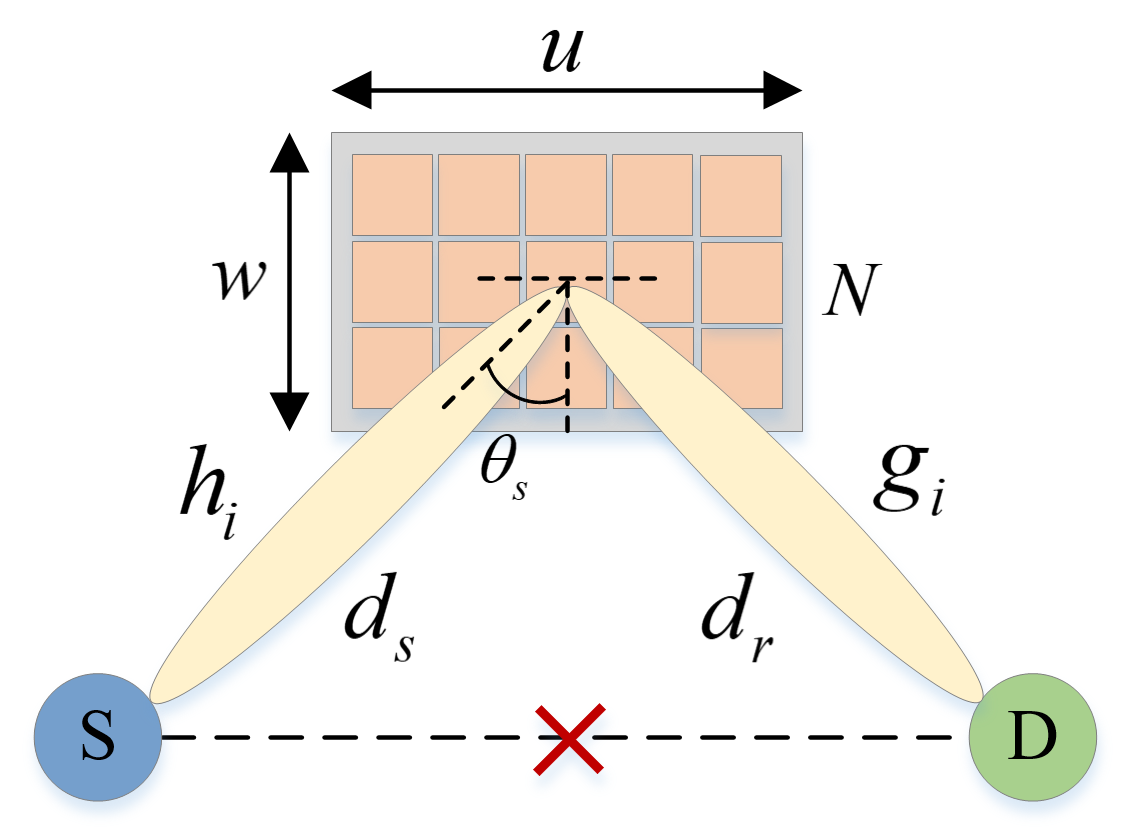}
  \caption{An IRS-aided communication system with $N$ metallic elements.}
 \end{figure}

\section{Coverage Probability Analysis}
In the section, we give the detailed process of deriving the coverage probability. Technically, we first determine the distribution of $A^2$ for $N=1$ and $N\ge1$. Then, the coverage probability can be obtained with the aid of cumulative probability distribution functions (CDFs) of $A^2$.

\subsection{Deriving the Exact Coverage Probability for $N=1$}
We first provide the PDF of $\alpha\beta$, and then derive its CDF.

\noindent \textbf{Lemma 1. } \emph{For two i.i.d. Rayleigh RVs $\alpha$ and $\beta$ with parameters of $\sigma_1$ and $\sigma_2$, respectively, the PDF of $\eta=\alpha\beta$ can be expressed as }
\begin{equation}
f_{\eta}(\eta)=\frac{\eta}{a^2}\mathcal{K}_0\left(\frac{\eta}{a}\right),
\end{equation}
\emph{where $a=\sigma_1\sigma_2$ and $\mathcal{K}_0(\cdot)$ is the zeroth order of modified Bessel function of the second kind. }

\emph{Proof: }  The readers can refer to \cite{b13} for the detailed derivation of the distribution of the product of two i.i.d. Rayleigh RVs based on Meijer's $G$-function.  $\hfill\blacksquare$

\noindent\textbf{Lemma 2.} \emph{For $N$=1, the CDF of $\eta$ can be obtained as}
\begin{equation}
F_{\eta}(\eta)=1-\frac{\eta}{a}\mathcal{K}_1(\frac{\eta}{a}),
\end{equation}
\emph{where $\mathcal{K}_1(\cdot)$ is the first order of modified Bessel function of the second kind. }

\emph{Proof: } See Appendix A.

We can further obtain the CDF of $\eta^2$ by $F_{\eta^2}(\eta)=F_{\eta}(\sqrt{\eta})$. Then, we can derive the exact coverage probability for $N=1$.

\noindent\textbf{Theorem 1.}
\emph{For $N=1$ and specific SNR threshold $\gamma_{\text{th}}$, the exact coverage probability can be represented as}
\begin{equation}
\begin{aligned}
P_{\text{cov}}(\gamma_{\text{th}})=\frac{d_s d_r}{\sigma_1\sigma_2}\sqrt{\frac{\gamma_{\text{th}}}{\bar{\gamma}}} \mathcal{K}_1 \left(\frac{d_s d_r}{\sigma_1\sigma_2}\sqrt{\frac{\gamma_{\text{th}}}{\bar{\gamma}}}\right).
\end{aligned}
\end{equation}

\emph{Proof: } The coverage probability is defined as the probability that the SNR is large than a specific threshold, which can be expressed as
\begin{equation}
\begin{aligned}
P_{\text{cov}}(\gamma_{\text{th}})&=Pr(\gamma \ge \gamma_{\text{th}})=1-Pr(\gamma \le \gamma_{\text{th}}),\\&=1-F_{\eta^2}\left(\frac{\gamma_{\text{th}}}{\bar{\gamma}}d_s^2 d_r^2\right).
\end{aligned}
\end{equation}
We can obtain (8) by combining (7) and (9). $\hfill\blacksquare$

\subsection{Approximating the Coverage Probability for Arbitrary $N$}
The derivation becomes more complicated and intractable for $N>1$. In practice, the PDF of $\eta$ follows $K$-distribution $K(b, \nu) $ with $b=a$ and $\nu=0$ \cite{b15}. Due to the complexity and intractability of $K$-distribution, we use Gamma distribution to approximate it, with the following explanations: (i) Gamma distribution is a Type-III Pearson distribution that is widely used in fitting distributions for positive RVs \cite{b16}. (ii) It has shown that $A$ is the sum of $K$-distributed $\eta_i$. Leveraging by the additive characteristic of Gamma distribution, the final distribution of $A$ can be easily obtained for the case of multiple elements\cite{b17}.

%(ii) There are many previous works that use a Gamma distribution to approximate $K$-distribution for the sake of tractability \cite{b16, b17}.

\noindent\textbf{Lemma 3.}  \emph{With the method of moment matching, the distribution of $\eta$ can be approximated as Gamma distribution,}
\begin{equation}
\eta \sim Ga(\frac{\pi^2}{16-\pi^2},\frac{16-\pi^2}{2\pi}\sigma_1\sigma_2),
\end{equation}
\emph{where $Ga(k,\theta)$ represents Gamma distribution, in which $\theta=\frac{16-\pi^2}{2\pi}\sigma_1\sigma_2$ is the scale parameter, $k=\frac{\pi^2}{16-\pi^2}$ is the shape parameter, and $\Gamma(\cdot)$ is the Gamma function.}
%\begin{equation}
%f_{\eta}(x,k,\theta)=\frac{x^{k-1}}{(\theta)^k\Gamma(k)}\exp(-\frac{x}{\theta}),
%\end{equation}

\emph{Proof:} See Appendix B.

The Kolmogorov-Smirnov (KS) test is widely used to examine the goodness of fit of two distributions, which enable us to assess the accuracy of approximation using the statistic $\mathcal{D}_{\text{KS}}$ that is defined as the maximum difference of two CDFs,
\begin{equation}
\mathcal{D}_{\text{KS}}=\max|F_{\text{Approx}}(t)-F_{\text{Actual}}(t)|.
\end{equation}
As shown in Fig.~2, we plot the CDF of the product of two Rayleigh RVs with different $\sigma$. The Gamma distribution in (10) is used to fit the actual CDF. The results present perfect agreements with very small $\mathcal{D}_{\text{KS}}$, thus confirming the accuracy of the approximation with Gamma distribution.

 \begin{figure}[t!]
  \centering
\includegraphics[width=2.5in]{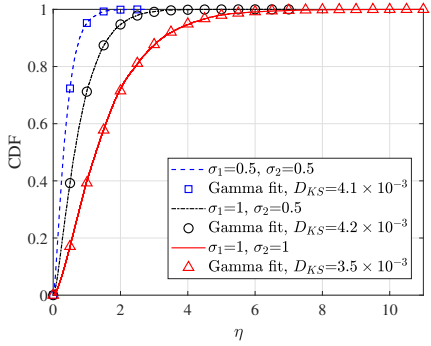}
      \caption{Actual CDFs of $\eta$ with Gamma fits for different fading coefficients.}
 \end{figure}

According to the addictive characteristic of Gamma distribution, we can obtain that $A \sim Ga(Nk,\theta)$ for $N$ elements in the IRS. Correspondingly, we can obtain that $A^2$ follows a generalized Gamma distribution with parameters of $p = \frac{1}{2}$, $d = \frac{Nk}{2}$, and $q=\theta^2$ \cite{b18}, whose CDF is expressed as
\begin{equation}
\begin{aligned}
F_{A^2}\left(z, \frac{1}{2}, \frac{Nk}{2}, \theta^2\right)&=\frac{\zeta(Nk,\frac{\sqrt{z}}{\theta})}{\Gamma(Nk)},
\end{aligned}
\end{equation}
where $\zeta(\cdot,\cdot)$ denotes the lower incomplete Gamma function.
%with $\zeta(s,x)=\int_0^{x}t^{s-1}e^{-t}dt$.

\noindent\textbf{Theorem 2.}\emph{ For arbitrary $N$, given the threshold $\gamma_{\text{th}}$, the general form of coverage probability can be represented as }
\begin{equation}
\begin{aligned}
P_{\text{cov}}(\gamma_{\text{th}})=\frac{\Gamma\left(Nk, s\right)}{\Gamma(Nk)},
\end{aligned}
\end{equation}
\emph{where $s=\frac{d_s d_r}{\theta} \left(\frac{\gamma_{\text{th}}}{\bar{\gamma}}\right)^{\frac{1}{2}}$ and $\Gamma(\cdot,\cdot)$ is the upper incomplete Gamma function.}
%with $\Gamma(s,x)=\int_x^{\infty}t^{s-1}e^{-t}dt$.}

\emph{Proof:}  Same as the proof of Theorem 1, the coverage probability is expressed as
\begin{equation}
\begin{aligned}
P_{\text{cov}}(\gamma_{\text{th}})&=1-F_{A^2}\left(\frac{\gamma_{\text{th}}}{\bar{\gamma}}d_s^2 d_r^2\right),\\
&=1-\frac{\zeta\left(Nk,\frac{2\pi d_s d_r}{(16-\pi^2)\sigma_1\sigma_2} \left(\frac{\gamma_{\text{th}}}{\bar{\gamma}}\right)^{\frac{1}{2}}\right)}{\Gamma(Nk)}.
\end{aligned}
\end{equation}
where the $\zeta(\cdot,\cdot)$  can be calculated by
\begin{equation}
\zeta\left(Nk, s\right)=\Gamma(Nk)-\Gamma\left(Nk, s\right).
\end{equation}
 Substituting (15) into (14), Theorem 2 can be proved. $\hfill\blacksquare$

The general expression shows that the size ($uw$), the number ($N$), the channel condition ($\sigma$), and the incident angle ($\theta_s$), all can impact the coverage probability. For intuitively validating and interpreting the derived result, we show the following remarks of \emph{\textbf{asymptotic analysis}}.

\noindent \textbf{Remark 1.} From the perspective of SNR threshold, it can be obtained that $P_{\text{cov}}^{s = 0}=1$ when $\gamma_{\text{th}}=0$ since $\Gamma(Nk)=\Gamma(Nk, 0)$ always holds. Moreover, we have $P_{\text{cov}}^{s\to  \infty}=0$ when $\gamma_{\text{th}}\to \infty$ since $\Gamma(Nk)=\zeta(Nk, \infty)$ always holds, which verifies the physical rationality of derived result.

\noindent \textbf{Remark 2.} From the perspective of the number of elements, since $Q(Nk, s)=\Gamma\left(Nk, s\right)/\Gamma(Nk)$ represents the CDF of Poisson distribution with parameter $s$, it directly has $Pr(X<0)=0$ and $Pr(X<\infty)=1$ for positive Poisson distributed RV $X$, which are equivalent to the cases of $N=0$ and $N\to\infty$, respectively. Thus, we can obtain that $P_{\text{cov}}^{N=0}=0$ and $P_{\text{cov}}^{N\to\infty}=1$, corresponding to two limits of no element and infinite elements, respectively.

\subsection{Finding the Optimal Number of Elements for IRS}
It has shown that the complete coverage probability ($P_{\text{cov}}=1$) achieves when $N$ is infinite. Practically, for a specific communication scenario with a fixed SNR requirement, the complete coverage can be guaranteed when $N$ reaches a certain number or above. Thus, it is critical to determine the optimal (\emph{minimum}) number, which can significantly reduce the cost of deployment. Fortunately, we can find the optimal $N^*$ by solving the equation of $\Gamma(Nk,s)=\Gamma(Nk)$ based on (13) with the numerical method, given by
\begin{equation}
N^*=\min\{N|\Gamma(Nk,s)=\Gamma(Nk)\}, \; N \in \mathbb{N}.
\end{equation}

\section{Simulation Results}
In this section, we present the simulation results to verify the derived expressions and investigate the impacts of many factors on coverage probability. The parameters are set as $P_s$=~1~mW, $G_s$=$G_r$= 0~dBi, $d_s$=$d_r$= 100~m, $\sigma_n^2$=~-90~dBm, and $uw=N l_e^2$ where $l_e$ is the side length of square element. Note that the remaining parameters may change for different purposes and all the simulated results are obtained by averaging over $10^5$ independent channel realizations.

 \begin{figure}[t!]
  \centering
\includegraphics[width=2.5in]{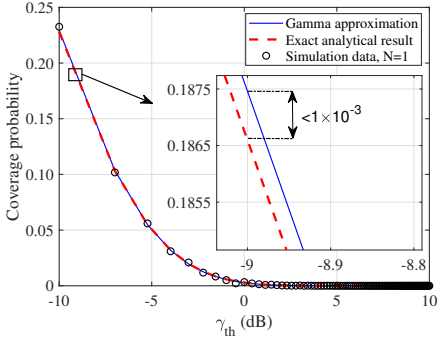}
  \caption{Results of exact analysis, Gamma approximation, and simulation.}
 \end{figure}

\begin{figure}[t!]
  \centering
\includegraphics[width=2.5in]{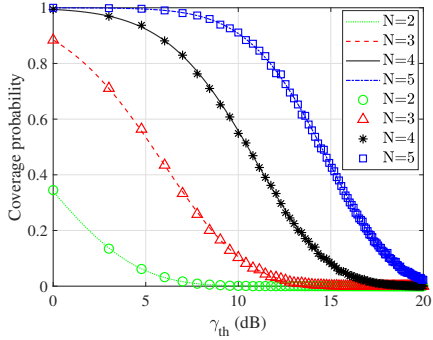}
  \caption{Gamma approximations (lines) versus simulations (marks).}
 \end{figure}

% \begin{figure}[htbp]
 % \centering
%  \subfigure[] {\includegraphics[width=1.7in]{gamma_th_cp_the1.eps}}
%  \subfigure[] {\includegraphics[width=1.7in]{uw_cp_the1.eps}}
%  \caption{The impact of tilt angle: (a) Comparison of coverage probability and (b) optimal tilt angle for UEs at different heights.}
%\end{figure}

 We first carry out the comparison among the exact form, the Gamma approximation, and the simulated result for $N=1$, as shown in Fig.~3. It shows that both the exact result and Gamma approximation are perfectly consistent with the simulation result, which reveals the correctness of our derivations. The tiny difference between the exact and approximated result is also exhibited in Fig.~3, where the difference is smaller than $10^{-3}$ for the SNR threshold of -9~dB, further suggesting the high accuracy of Gamma approximation.

 For $N>1$, we compare the simulated result with the Gamma approximation, as shown in Fig.~4. The results turn out perfect matching for diverse $N$. It is also found that the coverage probability can be greatly improved by increasing $N$. As an example, with $\gamma_{\text{th}}=10$~dB, the probability raises 0.45 by increasing $N$ from 3 to 4.

 As mentioned in Section III-C, for reducing the cost of deployment, we have defined the optimal number of IRS, of which the complete coverage can be achieved. Thus, we illustrate the coverage probability with respect to $N$, shown in Fig.~5. The results show that the complete coverage is realized for different SNR thresholds when $N$ reaches a certain value or above (where $l_e$ is fixed as $\frac{\lambda}{2}$). More specifically, the optimal number ($N^*$) is 8, 12, and 19 with the SNR thresholds of $10$~dB, $20$~dB, and $30$~dB, respectively. It is worth mentioning that the optimal numbers of simulation results are in accordance with those of numerical results in (16). The investigation can be used to evaluate the number of elements practically required in the deployment of the IRS.

  \begin{figure}[t!]
  \centering
\includegraphics[width=2.5in]{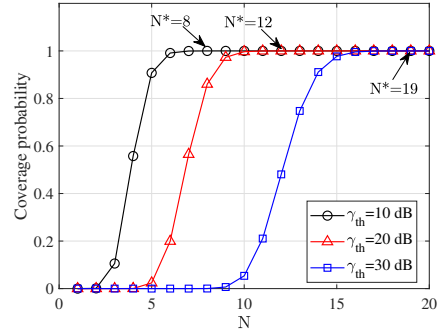}
  \caption{Coverage probability versus $N$ for different SNR thresholds.}
 \end{figure}

  \begin{figure}[t!]
  \centering
\includegraphics[width=2.5in]{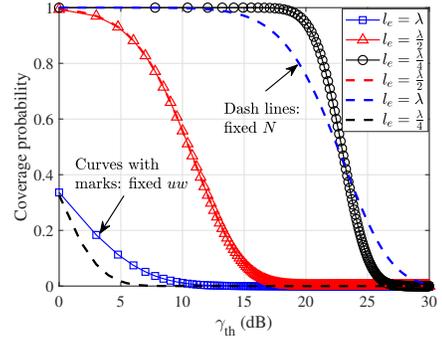}
  \caption{Coverage probability versus different sizes of elements.}
 \end{figure}

We also explore the impact of the size ($l_e$) of the element. We consider two conditions of fixed $N$ and $uw$, respectively. For fixed $N$ (= 4), the dash lines in Fig.~6 shows that larger size elements give rise to higher coverage probability for specific $\gamma_{\text{th}}$ since the total area of IRS becomes larger. For fixed $uw$ (curves with marks), smaller size elements result in higher probability because $N$ (= 1, 4, 16) increases.

Influences of channel fading coefficients and incident angles are also investigated. Interestingly, as shown in Fig.~7, the larger coefficient led by severe fading can contribute to the higher coverage probability, which is accountable for the constructive superposition of reflective waves. For incident angles, Fig.~8 shows that the smaller angle causes the larger coverage probability since the term of $\cos(\theta_s)$ measures the effective impinging area of the incident wave, where the upper and lower bounds of $\theta_s=0$ and $\theta_s=\frac{\pi}{2}$, correspond to the perpendicular and parallel incidences, respectively.

  \begin{figure}[t!]
  \centering
\includegraphics[width=2.5in]{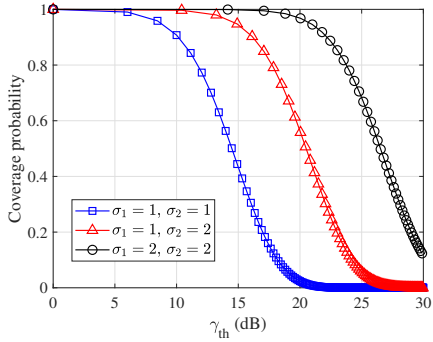}
  \caption{Coverage probability versus fading coefficients of channels.}
 \end{figure}

  \begin{figure}[!t]
  \centering
\includegraphics[width=2.5in]{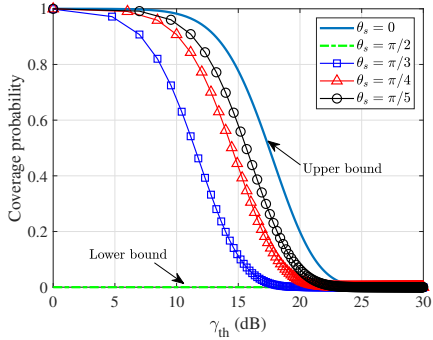}
  \caption{Coverage probability versus angles of incidence.}
 \end{figure}

\section{Conclusion}
In this work, we studied the coverage probability of IRS-aided communication systems. We derived the closed-form coverage probability for $N=1$. Further, with Gamma approximation, we provided a more general expression of coverage probability for an arbitrary number of $N$. We investigated the impacts of numerous parameters on the coverage probability. To be concluded, the larger number or area of the IRS can significantly improve the performance with the aid of controllable variables, including $N$, $l_e$, and $\theta_s$. Moreover, thanks to the constructive superposition of waves from each element, we found that uncontrollable channel fading coefficients also positively contribute to the coverage probability. The work not only derives a general expression of coverage probability but also provides many insights into the actual deployment of IRS.

% Moreover, the optimal number of IRS can be numerically determined to reduce the cost of deployment. , fortunately, that can be easily realized by the phase shift design of IRS

  \appendices
\setcounter{equation}{0}
\setcounter{subsection}{0}
\renewcommand{\theequation}{A.\arabic{equation}}
\renewcommand{\thesubsection}{A.\arabic{subsection}}
\section{Proof of Lemma 2}
The CDF of $\eta$ can be obtained based on (6),
\begin{equation}
\begin{aligned}
F_{\eta}(\eta)&=\int_{-\infty}^{\eta} f_{\eta}(x) dx=\int_0^{\eta} \frac{x}{a^2}\mathcal{K}_0\left(\frac{x}{a}\right) dx,\\
&=\lim_{x\to 0}\frac{x}{a}\mathcal{K}_1\left(\frac{x}{a}\right)-\frac{\eta}{a}\mathcal{K}_1(\frac{\eta}{a}).
\end{aligned}
\end{equation}
By using the law of Robida, we can seek the limit, given by
\begin{equation}
\begin{aligned}
\lim_{x\to 0}\frac{x}{a}\mathcal{K}_1\left(\frac{x}{a}\right)&=\lim_{x\to 0}\frac{\frac{-(\mathcal{K}_0(\frac{x}{a})+\mathcal{K}_2(\frac{x}{a}))}{2a}}{-\frac{a}{x^2}}.
\end{aligned}
\end{equation}
With [14, Eq. 8.446], when $x\to 0$, the series expansion is
\begin{equation}
-\frac{\mathcal{K}_0(\frac{x}{a})+\mathcal{K}_2(\frac{x}{a})}{2a} \to -\frac{a}{x^2}+\frac{2\log(\frac{1}{2a})+2\log(x)+2\epsilon+1}{4a}.
\end{equation}

By combining (A.1)-(A.3), we can have Lemma 2.
%&=\lim_{x\to 0}\frac{\frac{\partial K_1(x/a) }{\partial x}}{\frac{\partial (a/x) }{\partial x}}\\

\setcounter{equation}{0}
\setcounter{subsection}{0}
\renewcommand{\theequation}{B.\arabic{equation}}
\renewcommand{\thesubsection}{B.\arabic{subsection}}
\section{Proof of Lemma 3}
The $n$-th moment of Gamma distribution is expressed as
\begin{equation}
\mu_n^{G}=\theta^n\prod_{i=1}^{n}(k+i-1).
\end{equation}
while the moment of distribution of $\eta$ can be obtained with the aid of $\mathcal{K}_0(x/a)=\int_0^{\infty}\frac{\cos(xt/a)}{\sqrt{t^2+1}}dt$, given by
\begin{equation}
\mu_n^{\eta}=\int_0^{\infty}\int_0^{\infty}\frac{x^{n+1}}{a^2}\frac{\cos(\frac{xt}{a})}{\sqrt{t^2+1}} dt dx.
\end{equation}
By matching the first and second moment, i.e., $\mu_1^{\eta}=\frac{a\pi}{2}=k\theta=\mu_1^{G}$ and $\mu_2^{\eta}=4a^2=k(k+1)\theta^2=\mu_2^{G}$, the results of $k$ and $\theta$ can be obtained and one can have Lemma 3.

% or if you just want to reserve a space for a photo
%\vfil


\begin{thebibliography}{99}
\bibitem{b1}
Q. Wu and R. Zhang, ``Towards smart and reconfigurable environment:
intelligent reflecting surface aided wireless network,'' \emph{IEEE Commun.
Mag.}, vol. 58, no. 1, pp. 106–112, Jan. 2020.

\bibitem{b2}
M. Nemati, J. Park and J. Choi, ``RIS-assisted coverage enhancement in millimeter-wave cellular networks,'' \emph{IEEE Access}, vol. 8, pp. 188171-188185, Oct. 2020.

\bibitem{b3}
M. Cui, G. Zhang, and R. Zhang, ``Secure wireless communication via intelligent reflecting surface,'' \emph{IEEE Wireless Commun. Lett.}, vol. 8, no. 5, pp. 1410–1414, Oct. 2019.

\bibitem{b4}
Q. Wu and R. Zhang, ``Intelligent reflecting surface enhanced wireless
network via joint active and passive beamforming,'' \emph{IEEE Trans. Wireless
Commun.}, vol. 18, no. 11, pp. 5394–5409, Aug. 2019.

\bibitem{b11}
J. C. B. Garcia, A. Sibille and M. Kamoun, ``Reconfigurable intelligent surfaces: bridging the gap between scattering and reflection,'' \emph{IEEE J. Select. Areas Commun.}, vol. 38, no. 11, pp. 2538-2547, Nov. 2020.

\bibitem{b5}
S. Abeywickrama, R. Zhang, Q. Wu and C. Yuen, ``Intelligent reflecting surface: practical phase shift model and beamforming optimization,'' \emph{IEEE Trans. Commun.}, vol. 68, no. 9, pp. 5849-5863, Sept. 2020.

\bibitem{b6}
L. Yang, Y. Yang, M. O. Hasna and M. -S. Alouini, ``Coverage, probability of SNR gain, and DOR analysis of RIS-aided communication systems,'' \emph{IEEE Wireless Commun. Lett.}, vol. 9, no. 8, pp. 1268-1272, Aug. 2020.


\bibitem{b7}
E. Basar, M. Di Renzo, J. De Rosny, M. Debbah, M. S. Alouini, and R. Zhang, ``Wireless communications through reconfigurable intelligent
surfaces,'' \emph{IEEE Access}, vol. 7, pp. 116753–116773, 2019.


\bibitem{b8}
M. Badiu and J. P. Coon, ``Communication through a large reflecting
surface with phase errors,'' \emph{IEEE Wireless Commun. Lett.}, vol. 9, no. 2,
pp. 184–188, Feb. 2020.

\bibitem{b9}
W. Zhao, G. Wang, S. Atapattu, T. A. Tsiftsis and X. Ma, ``Performance analysis of large intelligent surface aided backscatter communication systems,'' \emph{IEEE Wireless Commun. Lett.}, vol. 9, no. 7, pp. 962-966, Jul. 2020.

\bibitem{b10}
W. Tang \emph{et al.}, ``Wireless communications with reconfigurable intelligent surface: path loss modeling and experimental measurement,'' \emph{IEEE Trans. Wireless Commun.}, 2020, Early Access Article.


\bibitem{b12}
{\"O}. {\"O}zdogan, E. Bj{\"o}rnson and E. G. Larsson, ``Intelligent reflecting surfaces: physics, propagation, and pathloss modeling,'' \emph{IEEE Wireless Commun. Lett.}, vol. 9, no. 5, pp. 581-585, May 2020.

\bibitem{b13}
J. Salo, H. M. El-Sallabi and P. Vainikainen, ``The distribution of the product of independent Rayleigh random variables,'' \emph{IEEE Trans. Antennas Propag.}, vol. 54, no. 2, pp. 639-643, Feb. 2006.

\bibitem{b14}
I. S. Gradshteyn and I. M. Ryzhik, \emph{Table of Integrals, Series and
Products}, 7th ed. Academic Press Inc, 2007.

\bibitem{b15}
E. Jakeman and P. N. Pusey, ``Significance of $K$ distributions in scattering experiments,'' \emph{Phys. Rev. Lett.}, 40, 546-550, Feb. 1978.

\bibitem{b16}
S. Al-Ahmadi and H. Yanikomeroglu, ``On the approximation of the generalized-$K$ distribution by a gamma distribution for modeling composite fading channels,'' \emph{IEEE Trans. Wireless Commun.}, vol. 9, no. 2, pp. 706-713, Feb. 2010.

\bibitem{b17}
S. Atapattu, C. Tellambura, and H. Jiang, ``A mixture gamma distribution
to model the SNR of wireless channels,'' \emph{IEEE Trans. Wireless Commun.}, vol. 10, no. 12, pp. 4193–4203, Dec. 2011.

\bibitem{b18}
P. G. Moschopoulos, ``The distribution of the sum of independent gamma random variables,'' \emph{Ann. Inst. Stat. Math.}, 37, 541–544, Dec. 1985.

\end{thebibliography}
\end{document}